\documentclass[journal]{IEEEtran}
\usepackage[utf8]{inputenc}
\usepackage{bm}
\usepackage{epsfig,amsfonts,amsbsy,bm,mathrsfs}
\usepackage{amssymb,amsmath,amsthm,latexsym,amscd,amsfonts}
\usepackage{lettrine}
\usepackage{authblk}
\usepackage{graphics}
\usepackage{graphicx}
\usepackage{psfrag,float}
\usepackage{pstricks}
\usepackage{pst-plot}
\usepackage{cite}
\usepackage{balance}
\usepackage{epsfig}
\usepackage{epstopdf}
\usepackage{bbm}
\usepackage{enumitem}
\usepackage{dsfont}
\usepackage{setspace}
\usepackage{float}
\usepackage{relsize}
\usepackage{comment}
\usepackage{subfigure} 
\usepackage{algorithm}
\usepackage[noend]{algpseudocode}
\usepackage[nolist]{acronym}
\usepackage{tabularx}
\usepackage{pifont}
\usepackage{units}

\usepackage{tikz}
\usetikzlibrary{calc}
\usepackage{xcolor}
\usepackage{tabularx}
\usepackage{colortbl}
\usepackage{pgfplots}
\pgfplotsset{compat=newest}
\usetikzlibrary{plotmarks}
\usetikzlibrary{arrows.meta}
\usepgfplotslibrary{patchplots}
\usepackage{grffile}
\newlength\fheight 
\newlength\fwidth 
\usepgfplotslibrary{fillbetween}
\newcommand{\Yu}[1]{\textcolor{magenta}{\footnotesize{\textsf {[Yu: #1]}}}}

\newcommand{\HW}[1]{\textcolor{blue}{[Henk: #1]}}

\begin{acronym}[ACRONYM]
\acro{AoA}{angle-of-arrival}
\acro{AoD}{angle-of-departure}
\acro{BS}{base station}
\acro{CDF}{cumulative density function}
\acro{CRB}{Cram\'er-Rao bound}
\acro{C-V2X}{cellular vehicle-to-anything}
\acro{DA}{data association}
\acro{DL}{downlink}
\acro{EKF}{extended Kalman filter}
\acro{FIM}{Fisher information matrix}
\acro{IP}{incidence point}
\acro{IQ}{in-phase and quadrature}
\acro{ISAC}{integrated sensing and communication}
\acro{JCS}{Joint Communication and Sensing}
\acro{LoS}{line-of-sight}
\acro{MIMO}{multiple-input multiple-output}
\acro{mmWave}{millimeter-wave}
\acro{NLoS}{non-line-of-sight}
\acro{OFDM}{orthogonal frequency-division multiplexing}
\acro{OTFS}{orthogonal time-frequency-space}
\acro{RIS}{reconfigurable intelligent surface}
\acro{RFS}{random finite set}
\acro{RMSE}{root mean squared error}
\acro{RTT}{round-trip-time}
\acro{SLAM}{simultaneous localization and mapping}
\acro{SNR}{signal-to-noise ratio}
\acro{ToA}{time-of-arrival}
\acro{TDoA}{time-difference-of-arrival}
\acro{Tx}{transmitter}
\acro{Rx}{receiver}
\acro{UE}{user equipment}
\acro{UL}{uplink}

\acrodef{PMBM}[PMBM]{Poisson  multi-Bernoulli  mixture}
\acrodef{PMB(M)}[PMB(M)]{Poisson  multi-Bernoulli  (mixture)}
\acrodef{PMB}[PMB]{Poisson  multi-Bernoulli}
\acrodef{KL}[KL]{Kullback–Leibler}

\acrodef{SP}[SP]{scattering point}
\acrodef{VA}[VA]{virtual anchor}
\acrodef{IP}[IP]{incidence point}

\acrodef{rfs}[RFS]{random finite set}
\acrodef{PPP}[PPP]{Poisson point process}
\acrodef{MBM}[MBM]{multi-Bernoulli  mixture}
\acrodef{MB}[MB]{multi-Bernoulli}
\acrodef{slam}[SLAM]{simultaneous localization and mapping}
\acrodef{ue}[UE]{user equipment}
\acrodef{URA}[URA]{uniform rectangular array}
\acrodef{QAM}[QAM]{quadrature amplitude modulation}
\acrodef{gospa}[GOSPA]{generalized optimal subpattern assignment}
\acrodef{EK}[EK]{extended Kalman}
\acrodef{EKF}[EKF]{extended Kalman filter}
\acrodef{GCI}[GCI]{generalized covariance intersection}

\acrodef{AA}[AA]{arithmetic average}

\end{acronym}

\begin{document}
\bibliographystyle{IEEEtran}
\bstctlcite{IEEEexample:BSTcontrol}

\title{Integrated Monostatic and Bistatic mmWave Sensing}
\author{\IEEEauthorblockN{
Yu Ge\IEEEauthorrefmark{1},
Hyowon Kim\IEEEauthorrefmark{2},
Lennart Svensson\IEEEauthorrefmark{1},
Henk Wymeersch\IEEEauthorrefmark{1},
Sumei Sun\IEEEauthorrefmark{3}
}

\IEEEauthorblockA{\IEEEauthorrefmark{1}
Department of Electrical Engineering, Chalmers University of Technology, Gothenburg, Sweden, }

\IEEEauthorblockA{\IEEEauthorrefmark{2}
Department of Electronics Engineering, Chungnam National University of Technology, Daejeon, South Korea, }

\IEEEauthorblockA{\IEEEauthorrefmark{3}
Institute for Infocomm Research, Agency for Science, Technology and Research, Singapore, }
\IEEEauthorblockA{ 
\{yuge,~lennart.svensson,~henkw\}@chalmers.se,~hyowon.kim@cnu.ac.kr,~sunsm@i2r.a-star.edu.sg}}

\maketitle

\pagestyle{empty}
\thispagestyle{empty}

\begin{abstract}
Millimeter-wave (mmWave) signals provide attractive opportunities for sensing due to their inherent geometrical connections to physical propagation channels. Two common modalities used in mmWave sensing are monostatic and bistatic sensing, which are usually considered separately. By integrating these two modalities, information can be shared between them, leading to improved sensing performance. In this paper, we investigate the integration of monostatic and bistatic sensing in a 5G mmWave scenario, implement the extended Kalman-Poisson multi-Bernoulli sequential filters to solve the sensing problems, and propose a method to periodically fuse user states and maps from two sensing modalities. 

\end{abstract}


\section{Introduction}
\Ac{ISAC} is expected to be one of the key features of 6G wireless systems \cite{An_ISAC_Survey2022}, and sensing using millimeter-wave (mmWave) signals is at the heart of \ac{ISAC}. Sensing is a broad term and covers everything from channel estimation and carrier sensing to device localization and environment awareness \cite{chaccour2022seven}. The term sensing, in this paper, is referred to the state estimation of the \ac{UE} and passive objects in the propagation environment, termed as \emph{positioning} and \emph{mapping}, respectively. 

\begin{figure}
\begin{centering}
\includegraphics[width=0.95\columnwidth]{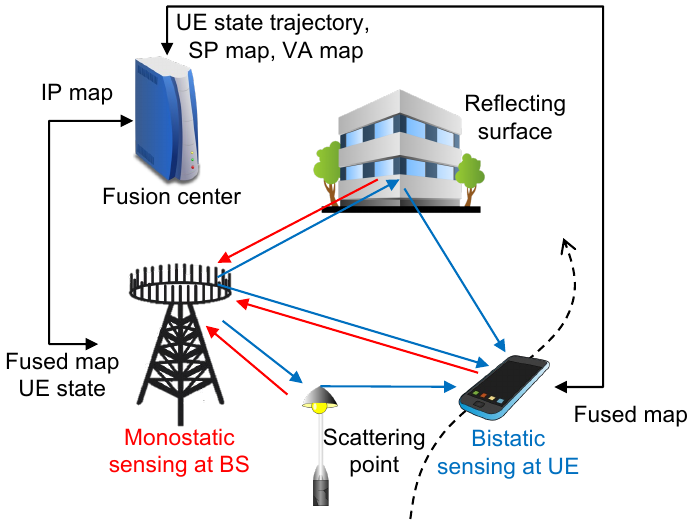}
	\caption{An example of monostatic sensing and bistatic sensing. In monostatic sensing, the \ac{BS} sends out signals (in blue), which go through the complex propagation environment and are received back again by the \ac{BS} (in red), which is then used for sensing. In bistatic sensing, the \ac{BS} sends signals to the receiver, and those signals go via a complex propagation environment and then reach the \ac{UE} (in blue), which is then used for sensing. Two sensing modalities are fused and overwritten by the fused map and \ac{UE} state periodically.
    }\vspace{-5mm}
	\label{Fig:monosta_Bista}
\end{centering}
\end{figure}

Monostatic sensing and bistatic sensing are the two most common sensing modalities 
\cite{ge2022mmwave}.  In monostatic sensing, the transmitter and receiver are co-located (or connected with fiber and act as distributed monostatic system), and thus share complete knowledge of the transmitted signals and the clock \cite{ge2022mmwave,crouse2014basic,stinco2012performance}. In bistatic sensing, on the other hand, the transmitter and the receiver are usually at different locations, where the receiver may only have partial knowledge of the transmitted signals and the synchronization problem between the transmitter and the receive needs to be considered \cite{ge2022mmwave,crouse2014basic,stinco2012performance}. Fig.~\ref{Fig:monosta_Bista} provides visualizations of both sensing modalities. Although the two modalities can be employed simultaneously, most existing works only consider one of them.

The related works can be divided into works that solve monostatic and bistatic mmWave sensing problems and works that integrate sensing modalities. The sensing problem (mapping and positioning in this paper) has been studied in many papers and addressed by different approaches  \cite{yassin2018mosaic,zhang2022toward,Erik_BPSLAM_TWC2019,EKPHD2021Ossi,kim2022ris,ge2022computationally}. Among these methods, \ac{rfs}-based methods \cite{EKPHD2021Ossi,kim2022ris,ge2022computationally} and certain message
passing-based methods \cite{Erik_BPSLAM_TWC2019} can  handle the inherent challenges of unknown \acp{DA}, the unknown number of objects/targets, misdetections, and clutter measurements, which all sensing applications with multiple objects/targets suffer from. 
Within these \ac{rfs}-based methods, \cite{ge2022computationally} proposed a \ac{PMB}-based algorithm that can keep both a good sensing performance and an acceptable computational complexity, making it suitable to be used in solving sensing problems. However,
only one sensing modality is considered in these works. The integration of sensing modalities is not a new idea and has been considered in earlier literature \cite{stinco2012performance,crouse2014basic,denove2021multiple,kim2022Sensing}, but they solely consider the integration of several monostatic sensors or several bistatic sensors, not the combination of monostatic and bistatic sensing. The only exception is \cite{yang2022hybrid}, which proposed  a message-passing-based algorithm that fused both sensing modalities in communication systems. However, \cite{yang2022hybrid}  assumes perfect synchronization between \ac{UE} and \ac{BS}, considers the \ac{UE} orientation to be known, and 
performs both monostatic and bistatic sensing at the \ac{UE} side, all of which make the problem considerably easier. 



In this paper, we study the integration of monostatic and bistatic sensing to improve the sensing performances and consider a scenario where monostatic  sensing is performed  by the \ac{BS} and bistatic sensing is performed at the \ac{UE}, and clock bias and the \ac{UE} orientation are estimated. The main contributions of this paper are summarized as follows: (\emph{i}) we develop an \ac{rfs}-based integration algorithm that can fuse the \ac{UE} states and maps from monostatic and bistatic sensing together, and provide the executable details on the proposed method; (\emph{ii}) we extend our previous work of  the \ac{EK}-\ac{PMB} \ac{SLAM} filter in \cite{ge2022computationally} by periodically replacing the corresponding updated maps and \ac{UE} states with the fused ones; (\emph{iii}) we validate the benefits of the integrating two sensing modalities through simulations in the mmWave radio network context, and show the integration can significantly
enhance sensing performances. 

\subsubsection*{Notations}
Scalars (e.g., $x$) are denoted in italics, vectors (e.g., $\boldsymbol{x}$) in bold lower-case letters, matrices (e.g., $\boldsymbol{X}$) in bold capital letters, and sets  (e.g., $\mathcal{X}$) in calligraphic. Transpose is denoted by $(\cdot)^{\mathsf{T}}$. A Gaussian density with mean $\boldsymbol{u}$ and covariance $\boldsymbol{\Sigma}$, evaluated at $\boldsymbol{x}$, is denoted by $\mathcal{N}(\boldsymbol{x};\boldsymbol{u},\boldsymbol{\Sigma})$.

\section{Models for bistatic and monostatic sensing}\label{sec:systemmodel}

    


We consider a scenario where monostatic and bistatic sensing happen at the same time, but the former is by the BS and the latter is by the UE. In this section, we introduce the state models, the received signal models with \ac{MIMO}-\ac{OFDM} signals, and the measurement models. 

\subsection{State Models} 

The dynamic state of the \ac{UE} at time step $k$, denoted as $\boldsymbol{s}_{k}$, consists of the 3D \ac{UE} position $\boldsymbol{x}_{\mathrm{UE},k}=[x_{k},y_{k},z_{k}]^{\mathsf{T}}$, the heading $\varpi_{k}$ and the clock bias $b_{k}$ (with respect to the BS). The \ac{UE}  evolves according to state dynamics, and the transition density is given by
\begin{equation}
f(\boldsymbol{s}_{k+1} | \boldsymbol{s}_{k}) = {\cal N}(\boldsymbol{s}_{k+1} ; \boldsymbol{v}(\boldsymbol{s}_{k}),\boldsymbol{Q}_{k+1}), \label{dynamicmodel}
\end{equation}
where $\boldsymbol{v}(\cdot)$ denotes the known transition function and $\boldsymbol{Q}_{k+1}$ is the process noise covariance. 
In the environment, three types of landmarks are considered, i.e., the \ac{BS}, \acp{SP}, and reflecting surfaces. The \ac{BS} is deployed with a \ac{URA}, and is parameterized by a location $\boldsymbol{x}_{\mathrm{BS}}\in \mathbb{R}^{3}$, which is a prior known. A \ac{SP}, corresponding to a small object, e.g., a street lamp, a traffic sign, etc, is parameterized by a location $\boldsymbol{x}_{\mathrm{SP}}\in \mathbb{R}^{3}$. A reflecting surface, corresponding to a large surface, e.g., wall, building facade, etc, is parameterized by a fixed \ac{VA} with location $\boldsymbol{x}_{\mathrm{VA}}\in \mathbb{R}^{3}$. The \ac{VA} is surface-specific, which is the reflection of the \ac{BS} with respect to the reflecting surface, given by \cite{palacios2019single}
\begin{align}
    \boldsymbol{x}_{\mathrm{VA}}=(\boldsymbol{I}-2\boldsymbol{\nu}\boldsymbol{\nu}^{\mathsf{T}}) \boldsymbol{x}_{\mathrm{BS}}+2\boldsymbol{\mu}^{\mathsf{T}}\boldsymbol{\nu} \boldsymbol{\nu},\label{VA}
\end{align}
where $\boldsymbol{\nu}$ is the normal to the surface, and $\boldsymbol{\mu}$ is an arbitrary point on the surface. 

There are some differences in how the state models are treated in bistatic and monostatic sensing:
\begin{itemize}
    \item \emph{Bistatic sensing:} In bistatic sensing, a VA always remains static, even though the \ac{IP} of the signal, where the signal hits the landmark, is moving while the \ac{UE} is moving.
    \item \emph{Monostatic sensing:} In monostatic sensing, the BS is not aware of the UE dynamics and can only determine the 3D position $\boldsymbol{x}_{\mathrm{UE},k}$. Therefore, the \ac{UE} is modeled as a random walk with the transition density
    \begin{equation}
f(\boldsymbol{x}_{\mathrm{UE},k+1} | \boldsymbol{x}_{\mathrm{UE},k}) = {\cal N}(\boldsymbol{x}_{\mathrm{UE},k+1} ; \boldsymbol{x}_{\mathrm{UE},k},\boldsymbol{\mathsf{Q}}_{k+1}), \label{dynamicmodel_mon}
\end{equation}
where $\boldsymbol{\mathsf{Q}}_{k+1}$ is the (large) process noise covariance. 
Moreover, as the \ac{BS} is always fixed, the \acp{IP} do not change over time, and there is no difference between  small objects and large surfaces. The environment is parameterized using \acp{IP} $\boldsymbol{x}_{\mathrm{IP}}\in \mathbb{R}^{3}$ for all landmarks. Please note that the \ac{SP} is the same as \ac{IP} for a small object, and the \ac{VA} can be calculated from the \ac{IP}
\vspace{-1mm}
\begin{align}
    \boldsymbol{x}_{\mathrm{VA}}=2\boldsymbol{x}_{\mathrm{IP}}-\boldsymbol{x}_{\mathrm{BS}}.\label{VA_from_IP}
\end{align}
\end{itemize}

\vspace{-5.5mm}

\subsection{Signal Models}

The \ac{BS} sends downlink \ac{OFDM} pilot signals to the \ac{UE} every time step. These signals are denoted by $\boldsymbol{f}_{\text{BS},g,k} x_{\kappa,g} $, for  time step $k$, transmission $g$, and subcarrier $\kappa$, in which $x_{\kappa,g}$ is the pilot signal and $\boldsymbol{f}_{\text{BS},g,k}$ is the precoder at BS. These signals are received by the UE for bistatic sensing and by the BS for monostatic sensing.  
\subsubsection{Bistatic Sensing}
The downlink signal can reach the \ac{UE} via the \ac{LoS} path or \ac{NLoS} paths, or both, leading to the following observation model \cite{heath2016overview}
\begin{align}
    {y}_{\kappa,g,k}&= \boldsymbol{w}_{\text{UE},g,k}^{\mathsf{H}}  {{\sum _{i=1}^{I_{k}}\rho_{k}^{i}\boldsymbol{a}_{\text{UE}}(\boldsymbol{\theta}_{k}^{i})\boldsymbol{a}_{\text{BS}}^{\mathsf{H}}(\boldsymbol{\phi}_{k}^{i})e^{-\jmath 2\pi \kappa \Delta_f \tau_{k}^{i}}}}  \boldsymbol{f}_{\text{BS},g,k} x_{\kappa,g} \nonumber\\& +  \boldsymbol{w}_{\text{UE},g,k}^{\mathsf{H}} \boldsymbol{n}_{\kappa,g,k},\label{eq:FreqObservationBiStatic} 
\end{align}
where ${y}_{\kappa,g,k}$ is the received signal, $\boldsymbol{n}_{\kappa,g,k}$ is the noise, $\boldsymbol{w}_{\text{UE},g,k}$ is the combining matrix at the UE side,  $\Delta_f$ is the subcarrier spacing, $\boldsymbol{a}_{\text{UE}}(\cdot)$ and $\boldsymbol{a}_{\text{BS}}(\cdot)$ are the steering vectors of the UE and the BS antenna arrays, respectively. There are $I_{k}$  visible landmarks for bistatic sensing, and we assume that each landmark creates only one path, thus, there are $I_{k}$ paths in total. Each path $i$ can be described by a complex gain $\rho_{k}^{i}$, a \ac{ToA} $\tau_{k}^{i}$, an \ac{AoA} pair $\boldsymbol{\theta}_{k}^{i}$ in azimuth and elevation, and an \ac{AoD} pair $\boldsymbol{\phi}_{k}^{i}$ in azimuth and elevation. The path parameters are related to the geometry, e.g., $\tau_{k}^{i}=\Vert \bm{x}_{\text{UE},k}-\bm{x}_{\text{BS}}\Vert/c+b_k$ for the \ac{LoS} path and $\tau_{k}^{i}=\Vert \bm{x}_{\text{UE},k}-\bm{x}_{\text{IP}}\Vert/c+\Vert \bm{x}_{\text{BS}}-\bm{x}_{\text{IP}}\Vert/c+b_k$ for any \ac{NLoS} path. 
The relations between \ac{AoA}/\ac{AoD}  and the geometric state can be found in \cite[App.~A]{ge20205GSLAM}. 

\subsubsection{Monostatic Sensing}
The downlink signal is  reflected by the reflecting surfaces or diffused by \acp{SP} and/or the passive \ac{UE} back to the \ac{BS}. Therefore, there is no \ac{LoS} path in monostatic sensing, and only \ac{NLoS} paths need to be considered. 
The observation becomes\footnote{Doppler measurements are not considered due to short transmission periods. However, the inclusion of Doppler measurements would greatly facilitate the detection and estimation of the mobile \ac{UE}.}
\begin{align}
    {r}_{\kappa,g,k}&=   \boldsymbol{w}_{\text{BS},g,k}^{\mathsf{H}}{{\sum _{i=1}^{L_{k}}\varrho_{k}^{i}\boldsymbol{a}_{\text{BS}}(\boldsymbol{\vartheta}_{k}^{i})\boldsymbol{a}_{\text{BS}}^{\mathsf{H}}(\boldsymbol{\vartheta}_{k}^{i})e^{-\jmath 2\pi \kappa \Delta_f \varepsilon_{k}^{i}}}}  \boldsymbol{f}_{\text{BS},g,k} x_{\kappa,g} \nonumber\\& +  \boldsymbol{w}_{\text{BS},g,k}^{\mathsf{H}}\boldsymbol{t}_{\kappa,g,k},\label{eq:FreqObservationMonoStatic} 
\end{align}
    where ${r}_{\kappa,g,k}$ is the received signal across the BS array, $\boldsymbol{t}_{\kappa,g,k}$ is the noise across the \ac{BS} array, $\boldsymbol{w}_{\text{BS},g,k}$ is the combining matrix at the BS side, $L_{k}$ is the number of landmarks (including the passive \ac{UE}) for monostatic sensing. Similar to bistatic sensing, each path $i$ can also be described by a complex gain $\varrho_{k}^{i}$, a \ac{ToA} $\varepsilon_{k}^{i}$, and an \ac{AoA} pair $\boldsymbol{\vartheta}_{k}^{i}$ (which is equal to the corresponding \ac{AoD}). Note that since the BS is synchronized with itself, $\varepsilon_{k}^{i}=2\Vert \bm{x}_{\text{IP}}-\bm{x}_{\text{BS}}\Vert/c$, for \acp{IP}. 
%


\subsection{Measurement Models}\label{sec:GeoModel}
The channel parameters of \ac{ToA}, \ac{AoA}, and \ac{AoD} of the two sensing modalities can be obtained by applying a channel estimator, e.g., \cite{richter2005estimation,venugopal2017channel,Gershman2010,jiang2021beamspace}, on \eqref{eq:FreqObservationBiStatic} and \eqref{eq:FreqObservationMonoStatic} at the \ac{UE} and \ac{BS} sides, respectively. However, the channel estimation is beyond the scope of this paper, and the channel parameters are already available to be utilized as measurements for bistatic and monostatic sensing purposes. Measurements provided by the channel estimator at time step $k$ are modeled as \acp{rfs}, given by $\mathcal{Z}_{k}^{\text{B}}=\{\boldsymbol{z}_{k}^{1},\dots, \boldsymbol{z}_{k}^{\hat{{I}}_{k}} \}$ and $\mathcal{Z}_{k}^{\text{M}}=\{\boldsymbol{\mathsf{z}}_{k}^{1},\dots, \boldsymbol{\mathsf{z}}_{k}^{\hat{{L}}_{k}} \}$  for bistatic and monostatic sensing, respectively. Please note that $\hat{{I}}_{k}\neq {I}_{k}$ and $\hat{{L}}_{k}\neq {L}_{k}$ in general, as some measurements can be clutter which are originated from noise peaks during channel estimation or transient objects, and landmarks might be misdetected. It is also important to notice that the \ac{DA} problem is unsolved, which means the source of each measurement is still unclear.
The likelihoods differ in monostatic and bistatic sensing, in the following way: 
\begin{itemize}
    \item \emph{Bistatic sensing:} For any $\boldsymbol{z}^{i} \in \mathcal{Z}_{k}^{\text{B}}$ originating from landmark with IP location $\boldsymbol{x}^{i}$, the likelihood function is modeled by 
    \begin{align}
 f(\boldsymbol{z}_{k}^{i}|\boldsymbol{x}^{i},\boldsymbol{s}_{k})=\mathcal{N}(\boldsymbol{z}_{k}^{i};\boldsymbol{h}(\boldsymbol{x}^{i},\boldsymbol{s}_{k}),\boldsymbol{R}_k^i),\label{pos_to_channelestimation}
\end{align}
where 
$\boldsymbol{h}(\boldsymbol{x}^{i},\boldsymbol{s}_{k})=[\tau_{k}^{i},(\boldsymbol{\theta}_{k}^{i})^{\mathsf{T}},(\boldsymbol{\phi}_{k}^{i})^{\mathsf{T}}]^{\mathsf{T}}$ represents the nonlinear function that transforms the geometric information to channel parameters, and $\boldsymbol{R}_k^i$ is the measurement covariance determined by the \ac{FIM} of channel parameters, as in \cite{Zohair_5GFIM_TWC2018}. 
    \item \emph{Monostatic sensing:} For any $\boldsymbol{\mathsf{z}}_{k}^{i} \in \mathcal{Z}_{k}^{\text{M}}$ originating from a landmark with IP location $\boldsymbol{x}^{i}$, the likelihood function is modeled by 
    \begin{equation}\label{eq:likelihood_function}
f(\boldsymbol{\mathsf{z}}_{k}^{i}|\boldsymbol{x}^{i},\boldsymbol{s}_{k})=\mathcal{N}(\boldsymbol{\mathsf{z}}_{k}^{i};\boldsymbol{\mathsf{h}}(\boldsymbol{x}^{i},\boldsymbol{x}_{\mathrm{BS}}),\boldsymbol{\mathsf{R}}_k^i),
\end{equation}
where $\boldsymbol{\mathsf{h}}(\boldsymbol{x}^{i},\boldsymbol{x}_{\mathrm{BS}})=[\epsilon_{k}^{i},(\boldsymbol{\vartheta}_{k}^{i})^{\mathsf{T}}]^{\mathsf{T}}$ denotes the corresponding nonlinear function, and $\boldsymbol{\mathsf{R}}_k^i$ denotes the measurement covariance. 
\end{itemize}

\section{Filters for Mapping and SLAM at BS and UE}

In this section, we describe the form of the filter used for mapping the environment and how it is combined with tracking the \ac{UE} state. The map is modeled as a \ac{PMB} \ac{rfs}.  In this section, the PMB density and the PMB  filters are briefly introduced. Details on the implementation of the filters are outside the scope of this paper, but can be found in \cite{williams2015marginal,garcia2018poisson,fatemi2017poisson,ge2022computationally}.

\subsection{Basics of PMB Density}

We assume a map $\mathcal{X}=\{\boldsymbol{x}^{1},\dots, \boldsymbol{x}^{|\mathcal{X}|}\}$ is a \ac{PMB} \ac{rfs}. A \ac{PMB} \ac{rfs} consists of two disjoint \acp{rfs}, a set $\mathcal{X}_{\mathrm{U}}$ of undetected objects, which are all landmarks that have never been detected before, and a set $\mathcal{X}_{\mathrm{D}}$ of detected objects, which are landmarks that have been detected at least once before \cite{garcia2018poisson}. We model $\mathcal{X}_{\mathrm{U}}$ as a \ac{PPP} and $\mathcal{X}_{\mathrm{D}}$ as a \ac{MB}, with the following densities
\begin{align}
    f_{\mathrm{P}}(\mathcal{X}_{\mathrm{U}})&=e^{-\int\lambda(\boldsymbol{x}')\mathrm{d}\boldsymbol{x}'}\prod_{\boldsymbol{x} \in \mathcal{X}_{\mathrm{U}} }\lambda(\boldsymbol{x}),\label{PPP} \\
     f_{\mathrm{MB}}(\mathcal{X}_{\mathrm{D}})&= \sum_{\mathcal{X}^{1}\biguplus \dots \biguplus \mathcal{X}^{|\mathcal{X}_{\mathrm{D}}|}=\mathcal{X}_{\mathrm{D}}}\prod_{i=1}^{|\mathcal{X}_{\mathrm{D}}|}f^{i}_{\mathrm{B}}(\mathcal{X}^{i}),\label{MBM}
\end{align}
where $\lambda(\boldsymbol{x})=\eta f_{\text{P}}(\boldsymbol{x})$ is the intensity function with $\eta$ denoting the mean of the Possion distribution and $f_{\text{P}}(\boldsymbol{x})$ denoting the spatial density; $\uplus$ is the union of mutually disjoint sets; $f_{\mathrm{B}}^{i}(\cdot)$ is the Bernoulli density of the $i$-th landmark, following
\begin{equation}
f^{i}_{\mathrm{B}}(\mathcal{X}^{i})=
\begin{cases}
1-r^{i} \quad& \mathcal{X}^{i}=\emptyset \\ r^{i}f^{i}(\boldsymbol{x}) \quad & \mathcal{X}^{i}=\{\boldsymbol{x}\} \\ 0 \quad & \mathrm{otherwise},
\end{cases}
\end{equation} 
where $r^{i} \in [0,1]$ is the existence probability, describing how likely the landmark exists, and $f^{i}(\cdot)$ is the corresponding spatial density. As $\mathcal{X}$ is the union of $\mathcal{X}_{\mathrm{U}}$ and $\mathcal{X}_{\mathrm{D}}$,  the density of $\mathcal{X}$ can be computed using the convolution formula \cite[eq. (4.17)]{mahler2014advances} as $f(\mathcal{X})=\sum_{\mathcal{X}_{\mathrm{U}}\biguplus\mathcal{X}_{\mathrm{D}}=\mathcal{X}}f_{\mathrm{P}}(\mathcal{X}_{\mathrm{U}})f_{\mathrm{MB}}(\mathcal{X}_{\mathrm{D}})$, 
which can be parameterized by its components, i.e., $\lambda(\boldsymbol{x})$ and $\{r^{i},f^{i}(\boldsymbol{x})\}_{i\in \mathbb{I}}$, with $\mathbb{I}$ representing the index set of $\mathcal{X}_{\mathrm{D}}$.

\subsection{PMB-based Filters for Bistatic and Monostatic  Sensing}
Two different \ac{PMB}-based filters are implemented independently. A \ac{PMB} \ac{SLAM} filter is run at the \ac{UE} side for bistatic sensing to localize the \ac{UE} as well as mapping the surrounding environment, and another \ac{PMB} filter is run at the \ac{BS} side for monostatic sensing to map the surrounding environment as well as the passive \ac{UE}.
\subsubsection{Bistatic Sensing}
At the \ac{UE} side, $\mathcal{Z}^{\text{B}}_{k}$ is taken as input and a \ac{PMB} \ac{SLAM} filter is run to both track the \ac{UE} and map the landmarks, i.e, \acp{VA} and \acp{SP}. We denote the map for bistatic sensing as $\mathcal{X}^{\text{B}}$. The goal of the \ac{PMB} \ac{SLAM} filter is to recursively compute the joint posterior $f(\boldsymbol{s}^{\text{B}}_{k+1},\mathcal{X}^{\text{B}}|\mathcal{Z}^{\text{B}}_{1:k+1})$ every time step, following the Bayesian filtering framework with \acp{rfs} \cite{ge2022computationally}
\begin{align}
     f(\boldsymbol{s}_{k+1},\mathcal{X}^{\text{B}}|\mathcal{Z}^{\text{B}}_{1:k+1}&) \propto   \ell(\mathcal{Z}^{\text{B}}_{k+1}|\boldsymbol{s}_{k+1},\mathcal{X}^{\text{B}}) f(\mathcal{X}^{\text{B}}|\mathcal{Z}^{\text{B}}_{1:k}) \nonumber \\ &\times \int f(\boldsymbol{s}_{k}|\mathcal{Z}^{\text{B}}_{1:k})f(\boldsymbol{s}_{k+1}|\boldsymbol{s}_{k}) \text{d} \boldsymbol{s}_{k}, \label{joint_posterior_bi}
\end{align}
where $\ell(\mathcal{Z}^{\text{B}}_{k+1}|\boldsymbol{s}_{k+1},\mathcal{X}^{\text{B}})$ denotes the RFS likelihood function of the measurement set of bistatic sensing, given by  \cite[eqs.\,(5)--(6)]{garcia2018poisson}. Instead of tracking the joint density, the marginal posteriors $f(\boldsymbol{s}_{k}|\mathcal{Z}^{\text{B}}_{1:k+1})$ and $f(\mathcal{X}^{\text{B}}|\mathcal{Z}^{\text{B}}_{1:k+1})$ are tracked by marginalizing out the map state and the \ac{UE} state in the joint posterior, respectively, described as 
\begin{align}
     f(\boldsymbol{s}_{k+1}|\mathcal{Z}^{\text{B}}_{1:k+1})  &= \int  f(\boldsymbol{s}_{k+1},\mathcal{X}^{\text{B}}|\mathcal{Z}^{\text{B}}_{1:k+1})\delta \mathcal{X}^{\text{B}},  \label{eq:UpMarObjMarg1}\\
    f(\mathcal{X}^{\text{B}}|\mathcal{Z}^{\text{B}}_{1:k+1}) &= \int f(\boldsymbol{s}_{k+1},\mathcal{X}^{\text{B}}|\mathcal{Z}^{\text{B}}_{1:k+1}) \mathrm{d}\boldsymbol{s}_{k+1},\label{eq:UpMarObjMarg2}
\end{align}
where $\int \psi(\mathcal{X})\delta \mathcal{X}$ denotes the set integral \cite[eq.~(4)]{williams2015marginal}. 
The \ac{EK}-PMB \ac{SLAM} filter proposed in \cite{ge2022computationally} is used to solve the \ac{SLAM} problem for bistatic sensing. Since there are different types of landmarks (VAs and SPs), a multi-model implementation of the filter is applied. 

\subsubsection{Monostatic Sensing}
At the \ac{BS} side, $\mathcal{Z}^{\text{M}}_{k}$ is taken as input and a \ac{PMB} filter is run to map the \acp{IP} as well as the passive \ac{UE}. We denote the map for monostatic sensing as $\mathcal{X}^{\text{M}}$. This  \ac{PMB} filter is to recursively compute the posterior $f(\mathcal{X}^{\text{M}}|\mathcal{Z}^{\text{M}}_{1:k+1})$, given by (for the IP map)
\begin{align}
f(\mathcal{X}^{\text{M}}|\mathcal{Z}^{\text{M}}_{1:k+1})& \propto   \ell(\mathcal{Z}^{\text{M}}_{k+1}|\mathcal{X}^{\text{M}}) f(\mathcal{X}^{\text{M}}|\mathcal{Z}^{\text{M}}_{1:k}), \label{joint_posterior_mo}
\end{align}
where $\ell(\mathcal{Z}^{\text{M}}_{k+1}|\mathcal{X}^{\text{M}})$ denotes the RFS likelihood function of the measurement set of monostatic sensing. Since the UE may also reflect energy during monostatic sensing, a multi-model implementation is used, where possible IPs are treated as static landmarks, while the possible UE follows a random walk model as \eqref{dynamicmodel_mon}, leading  to $f(\mathcal{X}_{k+1}^{\text{M}}|\mathcal{Z}^{\text{M}}_{1:k+1}) \propto   \ell(\mathcal{Z}^{\text{M}}_{k+1}|\mathcal{X}_{k+1}^{\text{M}}) f(\mathcal{X}_{k+1}^{\text{M}}|\mathcal{Z}^{\text{M}}_{1:k})$ after prediction.

\subsubsection{Output of the Filters}
{
Note that \eqref{joint_posterior_bi}--\eqref{joint_posterior_mo} are implemented by the prediction and update steps of the PMB components (please refer to \cite{ge2022computationally} for details). At the end of the two PMB filters, we have the following components: $f^{\text{B}}(\boldsymbol{s}_{k})$, $\lambda^{\text{B}}_{k}(\boldsymbol{x})$ and $\{r_{k}^{\text{B},i},\{w_{\zeta,k}^{\text{B},i},f_{\zeta,k}^{\text{B},i}(\boldsymbol{x})\}_{\zeta \in \{\text{VA}, \text{SP}\}}\}_{i\in \mathbb{I}^{\text{B}}_{k}}$ for bistatic sensing, with $w_{\zeta,k}^{\text{B},i}$ and $f_{\zeta,k}^{\text{B},i}(\boldsymbol{x})$ representing the probability of the landmark type being $\zeta \in \{\text{VA}, \text{SP}\}$ and its spatial density; $f_{k}^{\text{M}}(\boldsymbol{x}_{\mathrm{UE},k})$, $\lambda^{\text{M}}_{k}(\boldsymbol{x})$ and $\{r_{k}^{\text{M},i},f_{k}^{\text{M},i}(\boldsymbol{x})\}_{i\in \mathbb{I}^{\text{M}}_{k}}$ for monostatic sensing, where $f_{k}^{\text{M}}(\boldsymbol{x}_{\mathrm{UE},k})$ is picked up from $\mathcal{X}_{k}^{\text{M}}$ by taking the hard decision on the type and the rest are for $\mathcal{X}^{\text{M}}$.} 

\section{Fusion of Two Sensing Modalities}

The two filters (one at the BS and one at the UE) run in parallel, based on the downlink signals sent by the BS. Periodically, the separate maps will be fused {at the fusion center, which we assume is at the \ac{BS} side}, and the fused map is sent back to the BS and the UE. The operation of such fusion is described next.  We drop the condition on measurement sets in this section for notation simplicity and consider the two individual filters at time step $k$. At the UE, from bistatic sensing, we have the  \ac{UE} posterior, denoted as $f^{\text{B}}_{k}(\boldsymbol{s})$, and the PMB $f_k(\mathcal{X}^{\text{B}})$ with components $\lambda^{\text{B}}_{k}(\boldsymbol{x})$ and {$\{r_{k}^{\text{B},i},\{w_{\zeta,k}^{\text{B},i},f_{\zeta,k}^{\text{B},i}(\boldsymbol{x})\}_{\zeta \in \{\text{VA}, \text{SP}\}}\}_{i\in \mathbb{I}^{\text{B}}_{k}}$.} At the BS, from monostatic sensing, we have {$f_{k}^{\text{M}}(\boldsymbol{x}_{\mathrm{UE},k})$ and the PMB $f_k(\mathcal{X}^{\text{M}})$ with components}
$\lambda^{\text{M}}_{k}(\boldsymbol{x})$ and $\{r_{k}^{\text{M},i},f_{k}^{\text{M},i}(\boldsymbol{x})\}_{i\in \mathbb{I}^{\text{M}}_{k}}$.
This section introduces map fusion and how the UE state density can be treated. 


\subsection{Map Fusion}
To further reduce the notational burden, we will also omit the current time step $k$. Then,  given $f(\mathcal{X}^{\text{B}})$ and $f(\mathcal{X}^{\text{M}})$, we aim  to form a new map with \ac{PMB} density $f(\mathcal{X}^{\text{F}})$, which will be parameterized by $\lambda^{\text{F}}(\boldsymbol{x})$ and $\{r^{\text{F},i},f^{\text{F},i}(\boldsymbol{x})\}_{i\in \mathbb{I}^{\text{F}}}$. The process comprises two steps: map matching and component fusion. {Note that the VAs and SPs in $\mathcal{X}^{\text{B}}$ should be converted to IPs by modifying the corresponding densities, to be able to fuse with IPs in $\mathcal{X}^{\text{M}}$, where  $f_{\text{VA},k}^{\text{B},i}(\boldsymbol{x})$ changes according to \eqref{VA_from_IP} and $f_{\text{SP},k}^{\text{B},i}(\boldsymbol{x})$ is unchanged.}
\subsubsection{Map Matching}
The problem of matching Bernoullis can be cast as an optimal assignment problem \cite{blackman_1999}, given by 
\begin{align}
     \text{minimize}
     \quad & \mathrm{tr}(\boldsymbol{A}^\mathsf{T}\boldsymbol{C})\\
     \text{s.t.} \quad &a_{i,i'}\in\{0,1\},~\forall i,i',\nonumber\\
     &\sum\nolimits_i a_{i,i'} \le 1 
     ,~\forall i', \nonumber\\
     &\sum\nolimits_{i'}a_{i,i'} = 1,~\forall i, \nonumber
 \end{align}
     where $\boldsymbol{A} \in \mathbb{R}^{|\mathbb{I}^{\text{B}}| \times (|\mathbb{I}^{\text{M}}|+|\mathbb{I}^{\text{B}}|)}$ is the optimization variable and $\boldsymbol{C} \in \mathbb{R}^{|\mathbb{I}^{\text{B}}| \times (|\mathbb{I}^{\text{M}}|+|\mathbb{I}^{\text{B}}|)}$ depends on the two PMBs. We define the cost matrix $\boldsymbol{C}$ as
\begin{align}
&\boldsymbol{C} =\left[
\begin{matrix}
c( 1,1 )  & \ldots & c( 1,|\mathbb{I}^{\text{M}}|) \\ 
\vdots &  \ddots & \vdots \\
c( |\mathbb{I}^{\text{B}}|,1)   & \ldots & c( |\mathbb{I}^{\text{B}}|,|\mathbb{I}^{\text{M}}|) 
\end{matrix}
\left|
\,
\begin{matrix}
T_\mathrm{G}  & \ldots & \infty \\ 
\vdots &  \ddots & \vdots \\
\infty &  \ldots & T_\mathrm{G}
\end{matrix}
\right.
\right],\label{Dmatrix}
\end{align}
where $T_\mathrm{G}$ is the gating threshold for matching Bernoullis, and $c( i,i' )$ is the cost of matching Bernoulli $i$ in $\mathcal{X}^{\text{B}}$ with Bernoulli $i'$ in $\mathcal{X}^{\text{M}}$. {The cost is defined as 
\begin{align}
 c(i,i')   &= c(i,i',\zeta^*), \label{distance}
\end{align}
${\zeta^* \in \{\text{VA}, \text{SP}\}}$ and  $c(i,i',\zeta^*)$ is now defined by 
\begin{align}
 c(i,i',\zeta^*)   &=  \frac{1}{2w ^{\text{B},i}_{\zeta^*}} (\Vert \boldsymbol{\xi} ^{\text{B},i}_{\zeta^*}-\boldsymbol{\xi} ^{\text{M},i'}\Vert_{\boldsymbol{\Sigma} ^{\text{B},i}_{\zeta^*}}+ \Vert \boldsymbol{\xi} ^{\text{B},i}_{\zeta^*}-\boldsymbol{\xi} ^{\text{M},i'}\Vert_{\boldsymbol{\Sigma} ^{\text{M},i}}), \label{distance_1}
\end{align}
{with  $\boldsymbol{\xi}_{\zeta^*} ^{\text{B},i}$ and $\boldsymbol{\Sigma}_{\zeta^*} ^{\text{B},i}$ representing the mean and covariance of the conversed density from $f^{\text{B},i}_{\text{VA}}(\boldsymbol{x})$ to an IP density according to \eqref{VA_from_IP} or the density $f^{\text{B},i}_{\text{SP}}(\boldsymbol{x})$,}  $\boldsymbol{\xi} ^{\text{M},i}$ and $\boldsymbol{\Sigma} ^{\text{M},i}$ representing the mean and covariance of $f^{\text{M},i}(\boldsymbol{x})$, and $\Vert \boldsymbol{x}\Vert_{\boldsymbol{\Sigma}}=\boldsymbol{x}^{\text{T}}\boldsymbol{\Sigma}^{-1}\boldsymbol{x}$.} Correspondingly, we introduce the decision on whether the landmark for the association $(i,i')$ is a VA or a SP by 
\begin{align}
\zeta^*  = \arg \min_{\zeta} \{c(i,i',\zeta)\}_{\zeta \in \{\text{VA}, \text{SP}\}}.\label{hard_decision_type}
\end{align}
In \eqref{Dmatrix}, the left $|\mathbb{I}^{\text{B}}| \times |\mathbb{I}^{\text{M}}|$ sub-matrix corresponds to matches of Bernoullis in $\mathcal{X}^{\text{B}}$, the right $|\mathbb{I}^{\text{B}}| \times |\mathbb{I}^{\text{B}}|$ diagonal sub-matrix corresponds to mismatches of Bernoullis in $\mathcal{X}^{\text{B}}$. {Solving this optimal assignment problem is the main computational cost of the fusion process, but it is much less costly than the mapping and \ac{SLAM} filters, and} can be solved efficiently by the Auction algorithm \cite{blackman_1999}. {As a hard decision on the landmark type $\zeta$ is made in \eqref{distance} for each possible pair, the type of landmarks can be determined together with solving the optimal assignment problem.}  

\subsubsection{Component Fusion}
After map matching, there are several possibilities:
\begin{itemize}
    \item \emph{Match exists for $f^{\text{B},i}_{\zeta^*}(\boldsymbol{x})$:} In this case, $f^{\text{B},i}_{\zeta^*}(\boldsymbol{x})$  is matched to $f^{\text{M},i'}(\boldsymbol{x})$ for an $i'$. Appendix \ref{app:caseA} shows how to obtain the fused Bernoulli. The same procedure applies to any $f^{\text{M},i}(\boldsymbol{x})$ for which a match exists. {After fusion, the type of the corresponding landmark is determined. Therefore, that weight $w_{\zeta^*}^{\text{B},i}$ is set to 1, and another weight is set to 0 in $\mathcal{X}^{\text{B}}$ for the multi-model implementation. } 
    \item \emph{No match exists for $f^{\text{B},i}_{\zeta^*}(\boldsymbol{x})$:} In this case, $f^{\text{B},i}_{\zeta^*}(\boldsymbol{x})$ is not matched to any $f^{\text{M},i'}(\boldsymbol{x})$ for all $i'$. Appendix \ref{app:caseB} shows how to obtain a Bernoulli by fusing $f^{\text{B},i}_{\zeta^*}(\boldsymbol{x})$ with the PPP $\lambda^{\text{M}}(\boldsymbol{x})$, where a hard decision is made on $\zeta$ according to \eqref{hard_decision_type}. A similar procedure exists for any $f^{\text{M},i}(\boldsymbol{x})$ for which no match exists. 
\end{itemize}

After generating all Bernoullis in the fused map, we can re-index these Bernoullis, making $i\in \mathbb{I}^{\text{F}}$, so that $\mathcal{X^{\text{F}}}$ is a MB with $\{r^{\text{F},i},f^{\text{F},i}(\boldsymbol{x})\}_{i\in \mathbb{I}^{\text{F}}}$. 
The two PPPs are also fused, based on Appendix \ref{app:caseC}.  Finally, $\mathcal{X^{\text{B}}}$ and $\mathcal{X^{\text{M}}}$ are overwritten by $\mathcal{X^{\text{F}}}$. {Note that some small modifications should be made when overwriting  $\mathcal{X^{\text{B}}}$ by $\mathcal{X^{\text{F}}}$. The spatial density of a \ac{VA} should be recovered based on \eqref{VA_from_IP}. A fused Bernoulli which is generated by fusing $f^{\text{M},i'}(\boldsymbol{x})$ with  $\lambda^{\text{B}}$, should keep the multi-model representation with 50\% of being SP and 50\% of being VA in $\mathcal{X}^{\text{B}}$, as we are not sure the type of the corresponding landmark, and another spatial density can be generated by simply duplicating the density of the VA/SP according to \eqref{VA_from_IP}.}

\subsection{UE State Fusion}
At time step $k$, the \ac{UE} posterior $f^{\text{B}}_{k}(\boldsymbol{s})$ provided by the EK-PMB SLAM filter at the \ac{UE} side for bistatic sensing can be fused with the spatial density of the Bernoulli for the \ac{UE} provided by the EK-PMB filter at the BS side for monostatic sensing, given by
\begin{align}
    f_{k}^{\text{F}}(\boldsymbol{s}_k)&= \frac{f^{\text{B}}_{k}(\boldsymbol{s}_k)  f_{k}^{\text{M}}(\boldsymbol{x}_{\mathrm{UE},k}) }
    {\int f^{\text{B}}_{k}(\boldsymbol{s}_k')  f_{k}^{\text{M}}(\boldsymbol{x}'_{\mathrm{UE},k}) \text{d}\boldsymbol{s}_k'},
    \label{eq:UE_fused}
\end{align}
since the \ac{UE} serves as a passive object in monostatic sensing, which only provides information on the \ac{UE} position, i.e., $\boldsymbol{x}_{\mathrm{UE},k}=[\boldsymbol{s}_k]_{1:3}$. After fusion, $f^{\text{B}}_{k}(\boldsymbol{s}_k)$ and $f_{k}^{\text{M}}(\boldsymbol{x}_{\mathrm{UE},k})$ are overwritten by $f_{k}^{\text{F}}(\boldsymbol{s}_k)$ and $f_{k}^{\text{F}}(\boldsymbol{x}_{\mathrm{UE},k})$.

\section{Numerical Results}

\subsection{Simulation Environment}
We consider the  scenario introduced in \cite{EKPHD2021Ossi}, which is at 28 GHz and contains a single mmWave \ac{BS} with a known location, and four \acp{VA} and four \acp{SP} with unknown locations. Additionally, there is a \ac{UE} doing a counterclockwise constant turn-rate movement around the \ac{BS} on the x-y plane, and its state is unknown. The \ac{BS} and \ac{UE} are both equipped with an  $8\times8$ omnidirectional \ac{URA}. The \ac{BS} broadcasts \ac{OFDM} pilot signals with 16 symbols, 64 subcarriers, and 200~MHz bandwidth. {Random precoders and combiners are used in both modalities.} The transmitted power is set to 35~dBm. The noise figure at the BS is 20~dBm lower than the noise figure at the UE. The noise spectral density  is -174~dBm/Hz. Path loss is generated according to \cite[eq.~(45)]{Zohair_5GFIM_TWC2018}, with the reflection coefficient of reflecting surfaces and the radar cross-section of small objects as 0.7 and 50 $\text{m}^{2}$, respectively. The Fisher information matrix of channel parameters~\cite{Zohair_5GFIM_TWC2018} is  used as the covariance in \eqref{pos_to_channelestimation} and \eqref{eq:likelihood_function}. The UE, VAs, and SPs are always visible to the BS. The BS and VAs are always visible to the UE, while SPs are only visible to the UE within the field of view of the UE, which is set as 50 $\text{m}$. The detection probability of each visible path is set as 0.9. The gating threshold $T_\mathrm{G}$ is set as 25.

The EK-PMB (SLAM) filters with considering $10$ best \acp{DA} are implemented to solve the \ac{SLAM} and mapping problems for bistatic and monostatic sensing. To evaluate the benefits of the integration of bistatic and monostatic sensing, we run two simulations: individual bistatic and monostatic sensing without any fusion, and bistatic and monostatic sensing with periodic map and \ac{UE} state fusion. The simulations are run for one vehicle cycle, which is 40 time steps, and the fusion happens every 5 time steps. 
The mapping  performance is evaluated by the \ac{gospa} distance \cite{rahmathullah2017generalized} for both \acp{VA} and \acp{SP} for bistatic sensing, and for \acp{IP}, \acp{VA} and \acp{SP} for monostatic sensing. The positioning performance is quantified by the root mean squared error (RMSE). Overall, 100 Monte Carlo simulations are performed.

\subsection{Results}
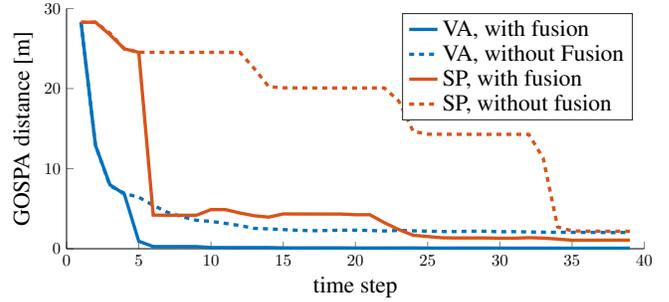
\begin{figure}
\center
\definecolor{mycolor1}{rgb}{0.00000,0.44700,0.74100}%
\definecolor{mycolor2}{rgb}{0.85000,0.32500,0.09800}%
\definecolor{mycolor3}{rgb}{0.00000,0.44700,0.74100}%
\definecolor{mycolor4}{rgb}{0.85000,0.32500,0.09800}%
\definecolor{mycolor5}{rgb}{0,0,0}%
%
\begin{tikzpicture}[scale=0.99\linewidth/14cm]

\begin{axis}[%
width=4.828in,
height=2.009in,
at={(1.011in,2.014in)},
scale only axis,
xmin=0,
xmax=40,
xlabel style={font=\color{white!15!black},font=\Large},
xlabel={time step},
ymin=0,
ymax=30,
ylabel style={font=\color{white!15!black},font=\Large},
ylabel={GOSPA distance [m]},
axis background/.style={fill=white},
axis x line*=bottom,
axis y line*=left,
legend style={legend cell align=left, align=left, draw=white!15!black,font=\Large}
]

\addplot [color=mycolor1,  line width=2.0pt]
  table[row sep=crcr]{%
1	28.2842712474619\\
2	12.9130181664571\\
3	7.98794742487066\\
4	6.82368150297796\\
5	0.935894998580982\\
6	0.26781166057354\\
7	0.267061734556661\\
8	0.271845004312497\\
9	0.265955170552944\\
10	0.148134048740071\\
11	0.14391596710371\\
12	0.142820276741109\\
13	0.145324628048946\\
14	0.144921636185657\\
15	0.106402926953698\\
16	0.102751328998707\\
17	0.105937340850927\\
18	0.108883716760961\\
19	0.10820628259636\\
20	0.0886378702269904\\
21	0.0902029631142111\\
22	0.0920001825437234\\
23	0.0947150635058746\\
24	0.0986324892241359\\
25	0.0795132031339013\\
26	0.0808794693448911\\
27	0.0839220746119728\\
28	0.0855659503558125\\
29	0.0859216688524059\\
30	0.0746171811419448\\
31	0.073443308214083\\
32	0.0730900409477491\\
33	0.0722095887739662\\
34	0.0722602453819765\\
35	0.0646778582986533\\
36	0.064306430763537\\
37	0.0650949413093915\\
38	0.0652217060246635\\
39	0.0658477314315378\\
};
\addlegendentry{VA, with fusion}

\addplot [color=mycolor1,dashed, line width=2.0pt]
  table[row sep=crcr]{%
1	28.2842712474619\\
2	12.9557918191039\\
3	7.86399675624512\\
4	6.91409623632088\\
5	6.42929137128071\\
6	5.4221602596274\\
7	4.53405591865621\\
8	3.99325661974692\\
9	3.54740272453291\\
10	3.41435324963098\\
11	3.1607449603252\\
12	2.89175526665508\\
13	2.54374423363597\\
14	2.4782772985238\\
15	2.38974893001974\\
16	2.30184020214872\\
17	2.24923672348448\\
18	2.27600155751725\\
19	2.31013456649227\\
20	2.31687215601418\\
21	2.27627511710839\\
22	2.21939972868339\\
23	2.28641353625345\\
24	2.23485689350741\\
25	2.19288591345611\\
26	2.14786070961041\\
27	2.14485061981658\\
28	2.18158976741853\\
29	2.18122654010497\\
30	2.12204247955248\\
31	2.13053055214978\\
32	2.09452027564075\\
33	2.06518537874312\\
34	2.05425877363699\\
35	2.04816271157934\\
36	2.04148088238539\\
37	2.03649431256224\\
38	2.01505879283771\\
39	2.03730338040763\\
};
\addlegendentry{VA, without Fusion}

\addplot [color=mycolor2,  line width=2.0pt]
  table[row sep=crcr]{%
1	28.2842712474619\\
2	28.2842712474619\\
3	26.7353241666934\\
4	24.9664004616816\\
5	24.5326079879531\\
6	4.19820914848031\\
7	4.18546205410392\\
8	4.18278511665928\\
9	4.18278511665928\\
10	4.88767516882975\\
11	4.88767516882975\\
12	4.46221784344813\\
13	4.11769698939067\\
14	3.94619079835299\\
15	4.34638781316323\\
16	4.33759695783087\\
17	4.33477006011216\\
18	4.33137702628375\\
19	4.33137702628375\\
20	4.24670180313645\\
21	4.24670180313645\\
22	3.25634263412433\\
23	2.39185373065367\\
24	1.68268679085899\\
25	1.52308599780446\\
26	1.36387638723321\\
27	1.3413150497649\\
28	1.34482946720912\\
29	1.34482946720912\\
30	1.3131428063024\\
31	1.3131428063024\\
32	1.3759214620927\\
33	1.32550164318426\\
34	1.20773879710901\\
35	1.05926932154996\\
36	1.06195243635217\\
37	1.06880886214444\\
38	1.07452236796641\\
39	1.07452236796641\\
};

\addlegendentry{SP, with fusion}

\addplot [color=mycolor2,dashed, line width=2.0pt]
  table[row sep=crcr]{%
1	28.2842712474619\\
2	28.2842712474619\\
3	26.9259361763827\\
4	24.9703452353212\\
5	24.537627620875\\
6	24.5262362474761\\
7	24.5227959187507\\
8	24.5241582224526\\
9	24.5241582224526\\
10	24.5241582224526\\
11	24.5241582224526\\
12	24.5241582224526\\
13	22.4911430706086\\
14	20.2426560763527\\
15	20.0954467685667\\
16	20.0855510025219\\
17	20.076868009157\\
18	20.0750689381031\\
19	20.0750689381031\\
20	20.0750689381031\\
21	20.0750689381031\\
22	20.0750689381031\\
23	18.4442028465473\\
24	14.6479142040698\\
25	14.3044115713028\\
26	14.2913720769721\\
27	14.2898896168105\\
28	14.2844629889575\\
29	14.2844629889575\\
30	14.2844629889575\\
31	14.2844629889575\\
32	14.2844629889575\\
33	11.314063048481\\
34	2.71252580996587\\
35	2.2330497924107\\
36	2.1932788102883\\
37	2.19042694189769\\
38	2.18271519309411\\
39	2.18271519309411\\
};
\addlegendentry{SP, without fusion}

\end{axis}
\end{tikzpicture}%
\caption{Comparison of mapping performances for VAs and SPs in bistatic sensing between two cases: with and without fusion.}
\label{Fig.mapping_bis_VA}
\end{figure}

\begin{figure}
\center
\definecolor{mycolor1}{rgb}{0.00000,0.44700,0.74100}%
\definecolor{mycolor2}{rgb}{0.85000,0.32500,0.09800}%
\definecolor{mycolor3}{rgb}{0.00000,0.44700,0.74100}%
\definecolor{mycolor4}{rgb}{0.85000,0.32500,0.09800}%
\definecolor{mycolor5}{rgb}{0,0,0}%
\definecolor{mycolor6}{rgb}{0.92900,0.69400,0.12500}%
%
\begin{tikzpicture}[scale=0.99\linewidth/14cm]

\begin{axis}[%
width=4.828in,
height=2.009in,
at={(1.011in,2.014in)},
scale only axis,
xmin=0,
xmax=40,
xlabel style={font=\color{white!15!black},font=\Large},
xlabel={time step},
ymin=0,
ymax=40,
ylabel style={font=\color{white!15!black},font=\Large},
ylabel={GOSPA distance [m]},
axis background/.style={fill=white},
axis x line*=bottom,
axis y line*=left,
legend style={legend cell align=left, align=left, draw=white!15!black,font=\Large}
]

\addplot [color=mycolor1,  line width=2.0pt]
  table[row sep=crcr]{%
1	40\\
2	13.901466101183\\
3	6.52325797227463\\
4	5.10740505422137\\
5	4.27005136560978\\
6	4.33278747958582\\
7	4.48806058542529\\
8	4.45986215086247\\
9	4.68436459261261\\
10	4.88842702550089\\
11	5.12263425590743\\
12	5.17034223300085\\
13	5.21194040011747\\
14	5.21288654061434\\
15	4.34685207886083\\
16	4.38254361097156\\
17	4.31420093943671\\
18	4.28465158192261\\
19	4.19404521571434\\
20	4.24712204366362\\
21	4.29962481692964\\
22	4.3064243816804\\
23	4.26284131223038\\
24	4.25080105562298\\
25	1.52386830550902\\
26	1.49784254495092\\
27	1.46596267710027\\
28	1.43483029588809\\
29	1.4211146768074\\
30	1.31393192358831\\
31	1.29704743615605\\
32	1.27879273836007\\
33	1.26869433517569\\
34	1.23589618547887\\
35	1.05995567638193\\
36	1.04535988672125\\
37	1.02863930055315\\
38	1.01396084717932\\
39	1.00743950963607\\
};
\addlegendentry{IP, With fusion}

\addplot [color=mycolor1, dashed, line width=2.0pt]
  table[row sep=crcr]{%
1	40\\
2	13.901466101183\\
3	6.52325797227463\\
4	5.10740505422137\\
5	4.74853732846519\\
6	4.93999746080846\\
7	5.17186312961942\\
8	5.10714019069183\\
9	5.29609110886188\\
10	5.4365364812555\\
11	5.48036005993712\\
12	5.46777390574595\\
13	5.43495849572843\\
14	5.41342348518348\\
15	5.40791054781811\\
16	5.39161905554999\\
17	5.38954646967995\\
18	5.36488604663722\\
19	5.275599864395\\
20	5.20215633286228\\
21	5.15442121207558\\
22	5.11996166087652\\
23	5.04027647769784\\
24	4.97540096248693\\
25	4.95876064232242\\
26	4.88304566887587\\
27	4.81527807471649\\
28	4.7794567535502\\
29	4.68186160083562\\
30	4.62522676243348\\
31	4.59386448514129\\
32	4.56033306967491\\
33	4.51600933862535\\
34	4.45246187091119\\
35	4.38096771870267\\
36	4.31293798393398\\
37	4.23761332946239\\
38	4.18298305989233\\
39	4.14658177623469\\
};
\addlegendentry{IP, Without fusion}

\addplot [color=mycolor2, line width=2.0pt]
  table[row sep=crcr]{%
1	28.2842712474619\\
2	28.2842712474619\\
3	28.2842712474619\\
4	28.2842712474619\\
5	0.250181240656512\\
6	0.197090817293126\\
7	0.175158041620915\\
8	0.146599515681165\\
9	0.13522663925897\\
10	0.15260347151224\\
11	0.138679960545385\\
12	0.127341126975928\\
13	0.118062733123869\\
14	0.10636155785618\\
15	0.112268373988681\\
16	0.106570302967091\\
17	0.103450210737629\\
18	0.0986046826791604\\
19	0.0957721256805274\\
20	0.0950384330443291\\
21	0.0917587602092845\\
22	0.0931833259556302\\
23	0.0895061712226694\\
24	0.0879855164329786\\
25	0.0889290445237292\\
26	0.0872689767364023\\
27	0.0877656876837928\\
28	0.0821412739415966\\
29	0.0800256470366935\\
30	0.0812943658997457\\
31	0.0780624436669342\\
32	0.0779995583914489\\
33	0.0763656947052844\\
34	0.0727200428380077\\
35	0.0720918493122059\\
36	0.0731404949814488\\
37	0.0711069681209956\\
38	0.0677111401109878\\
39	0.0667098680616997\\
};
\addlegendentry{VA, With fusion}

\addplot [color=mycolor2, dashed, line width=2.0pt]
  table[row sep=crcr]{%
1	28.2842712474619\\
2	28.2842712474619\\
3	28.2842712474619\\
4	28.2842712474619\\
5	28.2842712474619\\
6	28.2842712474619\\
7	28.2842712474619\\
8	28.2842712474619\\
9	28.2842712474619\\
10	28.2842712474619\\
11	28.2842712474619\\
12	28.2842712474619\\
13	28.2842712474619\\
14	28.2842712474619\\
15	28.2842712474619\\
16	28.2842712474619\\
17	28.2842712474619\\
18	28.2842712474619\\
19	28.2842712474619\\
20	28.2842712474619\\
21	28.2842712474619\\
22	28.2842712474619\\
23	28.2842712474619\\
24	28.2842712474619\\
25	28.2842712474619\\
26	28.2842712474619\\
27	28.2842712474619\\
28	28.2842712474619\\
29	28.2842712474619\\
30	28.2842712474619\\
31	28.2842712474619\\
32	28.2842712474619\\
33	28.2842712474619\\
34	28.2842712474619\\
35	28.2842712474619\\
36	28.2842712474619\\
37	28.2842712474619\\
38	28.2842712474619\\
39	28.2842712474619\\
};
\addlegendentry{VA, Without fusion}

\addplot [color=mycolor6, line width=2.0pt]
  table[row sep=crcr]{%
1	28.2842712474619\\
2	28.2842712474619\\
3	28.2842712474619\\
4	28.2842712474619\\
5	24.5276554975414\\
6	24.5179603285818\\
7	24.5136220796092\\
8	24.512426677598\\
9	24.5117148656289\\
10	4.85220521742515\\
11	5.09530231475834\\
12	5.14171261410124\\
13	5.18451121861305\\
14	5.1857794411858\\
15	4.32075443357312\\
16	4.35292242522215\\
17	4.28560285066297\\
18	4.25558978578495\\
19	4.16215416861236\\
20	4.22102539096338\\
21	4.27116456298334\\
22	4.28253702826393\\
23	4.23847436016908\\
24	4.22306063285976\\
25	1.47176399532447\\
26	1.44399161546812\\
27	1.41229038597812\\
28	1.38063695427895\\
29	1.3654195839841\\
30	1.25927220388811\\
31	1.24505380206038\\
32	1.22673206529663\\
33	1.21422670180755\\
34	1.18408447571007\\
35	1.00097146947067\\
36	0.98716276767783\\
37	0.970344763257246\\
38	0.95545951730065\\
39	0.950908656837823\\
};
\addlegendentry{SP, With fusion}

\addplot [color=mycolor6, dashed,line width=2.0pt]
  table[row sep=crcr]{%
1	28.2842712474619\\
2	28.2842712474619\\
3	28.2842712474619\\
4	28.2842712474619\\
5	28.2842712474619\\
6	28.2842712474619\\
7	28.2842712474619\\
8	28.2842712474619\\
9	28.2842712474619\\
10	28.2842712474619\\
11	28.2842712474619\\
12	28.2842712474619\\
13	28.2842712474619\\
14	28.2842712474619\\
15	28.2842712474619\\
16	28.2842712474619\\
17	28.2842712474619\\
18	28.2842712474619\\
19	28.2842712474619\\
20	28.2842712474619\\
21	28.2842712474619\\
22	28.2842712474619\\
23	28.2842712474619\\
24	28.2842712474619\\
25	28.2842712474619\\
26	28.2842712474619\\
27	28.2842712474619\\
28	28.2842712474619\\
29	28.2842712474619\\
30	28.2842712474619\\
31	28.2842712474619\\
32	28.2842712474619\\
33	28.2842712474619\\
34	28.2842712474619\\
35	28.2842712474619\\
36	28.2842712474619\\
37	28.2842712474619\\
38	28.2842712474619\\
39	28.2842712474619\\
};
\addlegendentry{SP, Without fusion}

\end{axis}
\end{tikzpicture}%
\caption{Comparison of mapping performances for IPs, VAs, and SPs in monostatic sensing between two cases: with and without fusion.}
\label{Fig.mapping_bis_IP}
\end{figure}
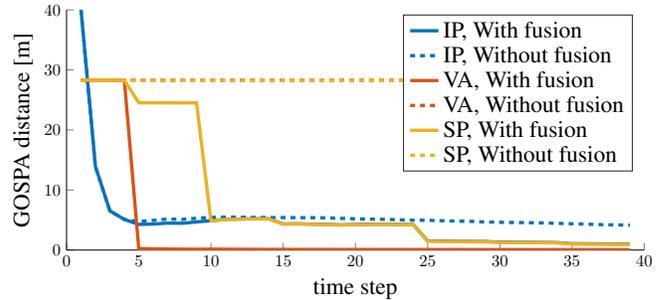

We first analyze how the integration of two sensing modalities affects the mapping performance. Fig.~\ref{Fig.mapping_bis_VA}  compares the \ac{gospa} distances between bistatic sensing with and without periodic fusion for \acp{VA} and  \acp{SP}, and Fig.~\ref{Fig.mapping_bis_IP} compares the \ac{gospa} distances between monostatic sensing with and without periodic fusion for \acp{IP}, \acp{VA} and  \acp{SP}. Both figures demonstrate that the EK-PMB (\ac{SLAM}) filter can map the landmarks in both sensing modalities and the mapping performance continuously increases with more measurements being received, as GOSPA distances for \acp{VA} and  \acp{SP} in Fig.~\ref{Fig.mapping_bis_VA} and \acp{IP} in Fig.~\ref{Fig.mapping_bis_IP} decrease over time. These figures also show that the type of landmarks can be distinguished accurately in bistatic sensing, which is due to the VAs and SPs generating measurements by following different observation models and the multi-model implementation of the \ac{EK}-PMB \ac{SLAM} filter proposed in \cite{ge2022computationally} is utilized to solve this problem. However, we cannot distinguish landmark types in monostatic sensing, as landmarks are treated all the same as \acp{IP}. Moreover, the periodic fusion of two sensing modalities improves the mapping performances in both cases, as solid lines are always lower than dashed lines in both figures, which benefits from using information from both modalities, i.e, the fusion considers measurements at the \ac{UE} side in bistatic sensing and measurements at the \ac{BS} side in monostatic sensing, and the better \ac{UE} positioning in bistatic sensing (see later). 
Especially, in bistatic sensing, \acp{SP} are detected sequentially due to the limited field of view of the \ac{UE} to \acp{SP}, and all \acp{SP} can be detected until time step 34. This is also the reason why the red dashed line in Fig.~\ref{Fig.mapping_bis_VA} drops step by step, as the misdetection of these unseen \acp{SP} brings a penalty to the GOPSA distance, which remains until all unseen \acp{SP} are seen and detected. However, \acp{SP} are always visible to the \ac{BS}, and the fusion with the monostatic sensing map introduce the unseen landmarks in bistatic sensing but detected in monostatic sensing to the bistatic sensing map, which makes the fused map detects all \acp{SP} after the first fusion at time step 5 {and distinguish the unseen SPs types one step later. Therefore, the red solid line in Fig.~\ref{Fig.mapping_bis_VA} dramatically decreases at time step 6.} The SP and VA GOSPA distances are provided merely to show that without fusion the BS has no way to distinguish the landmark type, leading to high GOSPA distances, but the fusion introduces the landmark types into the monostatic sensing map. 

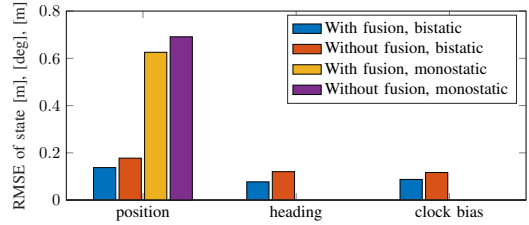
\begin{figure}
\center
%
%
\definecolor{mycolor1}{rgb}{0.00000,0.44700,0.74100}%
\definecolor{mycolor2}{rgb}{0.85000,0.32500,0.09800}%
\definecolor{mycolor3}{rgb}{0.92900,0.69400,0.12500}%
\definecolor{mycolor4}{rgb}{0.49400,0.18400,0.55600}
\definecolor{mycolor5}{rgb}{0,0,0}%
\begin{tikzpicture}[scale=0.99\linewidth/14cm]

\begin{axis}[%
width=3.842in,
height=1.581in,
at={(3.465in,2.378in)},
scale only axis,
bar shift auto,
xmin=0.5,
xmax=3.5,
xtick={1,2,3},
xticklabels={{position},{heading},{clock bias}},
ymin=0,
ymax=0.8,
ylabel style={font=\color{white!15!black}},
ylabel={RMSE of state [m], [deg], [m]},
axis background/.style={fill=white},
legend style={ anchor=north east, legend cell align=left, align=left, draw=white!15!black}
]

\addplot[ybar, bar width=0.145, fill=mycolor1, draw=black, area legend] table[row sep=crcr] {%
1	0.1375\\
2	0.0773\\
3	0.0876\\
};
\addplot[forget plot, color=white!15!black] table[row sep=crcr] {%
0.5	0\\
3.5	0\\
};
\addlegendentry{With fusion, bistatic}

\addplot[ybar, bar width=0.145, fill=mycolor2, draw=black, area legend] table[row sep=crcr] {%
1   0.1776\\
2   0.1204\\
3   0.1168\\
};
\addplot[forget plot, color=white!15!black] table[row sep=crcr] {%
0.5	0\\
3.5	0\\
};
\addlegendentry{Without fusion, bistatic }

\addplot[ybar, bar width=0.145, fill=mycolor3, draw=black, area legend] table[row sep=crcr] {%
1   0.6256\\
};
\addplot[forget plot, color=white!15!black] table[row sep=crcr] {%
0.5	0\\
3.5	0\\
};
\addlegendentry{With fusion, monostatic}

\addplot[ybar, bar width=0.145, fill=mycolor4, draw=black, area legend] table[row sep=crcr] {%
1   0.6911\\
};
\addplot[forget plot, color=white!15!black] table[row sep=crcr] {%
0.5	0\\
3.5	0\\
};
\addlegendentry{Without fusion, monostatic}

\end{axis}
\end{tikzpicture}%
\caption{Comparison of UE state estimation in bistatic and monostatic sensing between two cases: with and without fusion.}
\label{Fig.positioning_bis}
\end{figure}

Finally, we validate the benefits of the integration of two sensing modalities in positioning performance. Fig.~\ref{Fig.positioning_bis} displays the RMSEs of the estimated \ac{UE} position, heading, and bias for both sensing modalities with and without fusion. From the figure, we observe that  monostatic sensing can only estimate the \ac{UE} position, and cannot estimate its heading and clock bias, as the \ac{UE} serves as a passive object in monostatic sensing, while \ac{UE} position, heading, and clock bias can be estimated in bistatic sensing, as channel parameters depend on all three terms. Bistatic sensing provides roughly four times better position estimates than monostatic sensing. This big gap is caused by the consideration of the \ac{UE} movement model as well as the benefit of the mapping in bistatic sensing. However, monostatic sensing does not have a good \ac{UE} movement model and all the rest measurements from other landmarks cannot contribute to \ac{UE} positioning, as they do not contain any information on the \ac{UE} state. Fig.~\ref{Fig.positioning_bis} also displays that the periodic fusion can further improve the positioning performance in bistatic sensing, as all blues bars are lower than the red bars, indicating better position, heading and clack bias estimations can be acquired. The reason is that a better map can be obtained by the periodic fusion of two maps and measurements from the passive \ac{UE} in monostatic sensing are also used, which benefit the \ac{UE} positioning. Please note that the better \ac{UE} positioning can also enhance the mapping performance in turn. With periodic fusion, monostatic sensing can also obtain  better \ac{UE} position estimates. This is because the estimation error drops when the fusion happens and is more or less the same as before for the rest time steps. 

\section{Conclusions}

In this paper, we address the SLAM problem in bistatic sensing and the mapping problem in monostatic sensing using EK-PMB (SLAM) filters. A \ac{rfs}-based algorithm for integrating monostatic and bistatic sensing is first introduced in this paper, and executable details on the fusion of maps and \ac{UE} states are also provided. Via simulations, which use realistic mmWave signal parameters, we demonstrate that the implementation of the EK-PMB (SLAM) filters can map the environment and position the \ac{UE} state simultaneously in bistatic sensing, and map the environment as well as the passive \ac{UE} in monostatic sensing. The results also indicate that periodic fusion of monostatic and bistatic sensing helps the filters to acquire better mapping and SLAM performances in monostatic and bistatic sensing, respectively.

\section*{Acknowledgments}
This work was supported, in part, by the Wallenberg AI, Autonomous Systems and Software Program (WASP) funded by Knut and Alice Wallenberg Foundation,  by Hexa-X-II, part of the European Union’s Horizon Europe research and innovation programme under Grant Agreement No 101095759, and by the Basic Science Research Program through the National Research Foundation of Korea (2022R1A6A3A03068510).

\vspace{0mm}

\appendix[Map Fusion]
\label{app}
\subsection{Fusion of Two Bernoullis}
 \label{app:caseA}
By following the \ac{GCI} approach, the fusion of two Bernoullis $\{r^{\text{B}},f^{\text{B}}(\boldsymbol{x})\}$ and $\{r^{\text{M}},f^{\text{M}}(\boldsymbol{x})\}$  results in a Bernoulli with parameters \cite{mahler2000optimal,Batistelli_GCI_JSTSP2013,frohle2020decentralized}
\allowdisplaybreaks
\begin{align}
    &r^{\text{F}} = \frac{C(r^{\text{B}})^{\alpha}(r^{\text{M}})^{\beta}}{C(r^{\text{B}})^{\alpha}(r^{\text{M}})^{\beta}+(1-r^{\text{B}})^{\alpha}(1-r^{\text{M}})^{\beta}},\\
    &f^{\text{F}}(\boldsymbol{x})=\frac{ f^{\text{B}}(\boldsymbol{x})^{\alpha}f^{\text{M}}(\boldsymbol{x})^{\beta}}{
    C},
    \label{eq:GCI_B}
\end{align}
where $C=\int f^{\text{B}}(\boldsymbol{x})^{\alpha}f^{\text{M}}(\boldsymbol{x})^{\beta}\text{d}(\boldsymbol{x})$, $\alpha$ and $\beta$ are the fusion weights, satisfying  $\alpha+\beta=1$, with $\alpha= r^{\text{B}}/(r^{\text{B}} + r^{\text{M}})$, and $\beta= r^{\text{M}}/(r^{\text{B}} + r^{\text{M}})$. 
\vspace{0mm}
\subsection{Fusion of a Bernoullis with a PPP} \label{app:caseB}
The fusion of a Bernoulli $\{r^{\text{B}},f^{\text{B}}(\boldsymbol{x})\}$ and a PPP $\lambda^{\text{M}}(\boldsymbol{x})=\eta^{\text{M}}f_{\text{P}}^{\text{M}}$ results in a Bernoulli with parameters \cite{frohle2020decentralized}
\begin{align}
    &r^{\text{F}} = \frac{C(r^{\text{B}})^{\alpha}(\eta^{\text{M}})^{\beta}}{C(r^{\text{B}})^{\alpha}(\eta^{\text{M}})^{\beta}+(1-r^{\text{B}})^{\alpha}},\\
    &f^{\text{F}}(\boldsymbol{x})= \frac{f^{\text{B}}(\boldsymbol{x})^{\alpha}f_{\text{P}}^{\text{M}}(\boldsymbol{x})^{\beta}
    }{C},
    \label{eq:GCI_PPP_B}
\end{align}
with $C=\int f^{\text{B}}(\boldsymbol{x})^{\alpha}f_{\text{P}}^{\text{M}}(\boldsymbol{x})^{\beta}\text{d}(\boldsymbol{x})$, $\alpha = r^{\text{B}}/(r^{\text{B}} + \eta^{\text{M}})$, and $\beta= r^{\text{M}}/(r^{\text{B}} + \eta^{\text{M}})$.

\subsection{Fusion of two PPPs} \label{app:caseC}
Fusion of two PPPs $\lambda^{\text{B}}(\boldsymbol{x})=\eta^{\text{B}}f_{\text{P}}^{\text{B}}$ and $\lambda^{\text{M}}(\boldsymbol{x})=\eta^{\text{M}}f_{\text{P}}^{\text{M}}$ results in a new PPP with parameters \cite{frohle2020decentralized}
\begin{align}
    &\lambda^{\text{F}}(\boldsymbol{x}) = \lambda^{\text{B}}(\boldsymbol{x}) ^{\alpha}\lambda^{\text{M}}(\boldsymbol{x})^{\beta}=\eta^{\text{F}}f_{\text{P}}^{\text{F}}(\boldsymbol{x}),\\
    &\eta^{\text{F}}= C(\eta^{\text{B}})^{\alpha}(\eta^{\text{M}})^{\beta},\\
    &f_{\text{P}}^{\text{F}}(\boldsymbol{x}) =  \frac{f_{\text{P}}^{\text{B}}(\boldsymbol{x})^{\alpha}f_{\text{P}}^{\text{M}}(\boldsymbol{x})^{\beta}}
    {C},
    \label{eq:GCI_PPP}
\end{align}
where $C=\int f_{\text{P}}^{\text{B}}(\boldsymbol{x})^{\alpha}f_{\text{P}}^{\text{M}}(\boldsymbol{x})^{\beta}\text{d}(\boldsymbol{x})$, and $\alpha=\beta=1/2$.

\balance

\bibliography{references}

\end{document}